\documentclass{IEEEcsmag}

\usepackage[colorlinks,urlcolor=blue,linkcolor=blue,citecolor=blue]{hyperref}

\usepackage{upmath}
\usepackage{booktabs}

\let\labelindent\relax

\usepackage{enumitem}
\usepackage{makecell}
\usepackage{microtype}
\usepackage{multicol}
\usepackage{balance}

\usepackage[colorinlistoftodos,prependcaption]{todonotes}

\jvol{}
\jnum{}
\paper{}
\jmonth{}
\jname{}
\pubyear{2022}

\begin{document}


\title{\huge \textbf{Perspectives on Negative Research Results in Pervasive Computing}}

\author{Ella Peltonen} 
\affil{University of Oulu, Finland}

\author{Nitinder Mohan}
\affil{Technical University Munich, Germany}

\author{Peter Zdankin}
\affil{University of Duisburg-Essen, Germany}

\author{Tanya Shreedhar}
\affil{IIIT-Delhi, India}

\author{Tri Nguyen}
\affil{University of Oulu, Finland} 

\author{Suzan Bayhan}
\affil{University of Twente, The Netherlands}

\author{Jon Crowcroft}
\affil{University of Cambridge, United Kingdom}

\author{Jussi Kangasharju}
\affil{University of Helsinki, Finland}

\author{Daniela Nicklas}
\affil{University of Bamberg, Germany}


\begin{abstract}
Not all research leads to fruitful results; trying new ways or methods may surpass the state of the art, but sometimes the hypothesis is not proven or the improvement is insignificant. In a systems discipline like pervasive computing, there are many sources of errors, from hardware issues over communication channels to heterogeneous software environments. However, failure to succeed is not a failure to progress. It is essential to create platforms for sharing insights, experiences, and lessons learned when conducting research in pervasive computing so that the same mistakes are not repeated. And sometimes, a problem is a symptom of discovering new research challenges. Based on the collective input of the First International Workshop on Negative Results in Pervasive Computing~(PerFail 2022), co-located with the 20th International Conference on Pervasive Computing and Communications~(PerCom 2022), this paper presents a comprehensive discussion on perspectives on publishing negative results and lessons learned in pervasive computing.

\end{abstract}

\maketitle

\clearpage




\chapterinitial{Research} in fields such as pervasive computing, networking, systems, etc., is usually driven by results that push forward (or improve efficiency of) the state-of-the-art or provide demonstrable contributions. To achieve novel and practical solutions, researchers often invest significant efforts into designing and evaluating several iterations of design choices with hopes of achieving the desired performance. However, the probability of such efforts leading to a negative output can be quite high due to a lack of complete understanding/hindsight or unconvincing design choices, some of which may be harder to rectify at later stages. Such efforts are rarely rewarded since negative results are usually hard to publish~\cite{matosin2014negativity,fanelli2012negative}, being considered not novel or lacking significant new knowledge. However, we argue that negative results, if properly leveraged, can be quite beneficial to the research community at-large in the perspective of lessons learned and knowledge transfer between the research groups tackling the same topics. While our arguments in this article are from the perspective of pervasive computing research, many of our insights are valid in other fields of computing, such as human-computer interaction, network/mobility measurements, security and privacy, etc., thanks to the inherently interdisciplinary nature of the pervasive computing.

Most pervasive computing systems require multiple, often non-standard components, like hardware prototypes for sensing, developing computation paradigms, novel energy-saving communication protocols, adaptive middleware, or multi-modal user interfaces. Often, a new research idea or hypothesis covers only one part of the system, but to complete the experiment, you must (at least partially) get everything to work. Each component might be a source of error or the bottleneck that prohibits the expected result. However, sometimes the interplay of different components highlight the symptom of an underlying, more significant challenge that could lead to new research ideas. Hence, in this article, we assert that we \emph{must} treat negative results as a potential source for new scientific insights. 

While negative results might not directly contribute to the advancement of the state-of-the-art, the wisdom of hindsight could be an essential contribution in itself, such that other researchers could avoid falling into similar pitfalls. Such ``what not to do?" insights can also foster good research practices, especially for handling gray areas like ethical boundaries~\cite{mlinaric2017dealing}. We consider negative results to be outcomes of studies that are run correctly (in the light of the current state-of-the-art) and in good practice but fail in proving the hypothesis (statistically significantly). The “badness” of the work might also emerge from properly but unfittingly designed data collection or (non-trivial) lapses of hindsight, especially in studies involving real-world uncontrolled measurements. Furthermore, the experiences that help spot negative outcomes from intermediate results benefit the systems and networking community at large.  

Besides, it is crucial to share (especially for junior researchers) the knowledge that not all experiments will be successful, failures might happen, and knowing how to overcome unexpected outcomes is a skill to develop as a researcher. 
The interest in this field was evident from the participation in the questionnaire for Percom 2022 participants, in which $\approx$70 attendees participated. The result of the questionnaire revealed that $\approx$85\% of the conference participants already failed in their research work. 
The perspectives noted in this article are drawn from discussions with the pervasive computing community during the First International Workshop on Negative Results in Pervasive Computing (PerFail 2022)\footnote{\url{https://perfail-workshop.github.io/2022/}} and from authors' past experiences. Specifically, this paper discusses failure taxonomy in pervasive computing, considers mitigation and avoidance strategies, and presents our call-for-arms for healthy failure culture in pervasive research.

\section{What is a Failure?} 

Traditional applied research aims to fundamentally understand how systems work or how to make them work. Within the pervasive computing, systems and networking research, solutions that \emph{demonstrably} work are naturally more attractive to the community. Therefore, it becomes essential to gauge the usefulness of research that failed to work. Failure can be viewed as a necessary catalyst to research -- if there is no possibility to fail, the novelty of the research is also questionable. 

Within the research ecosystem, the definition of ``failure'' is often misconstrued, which makes many researchers incorrectly label unexpected outcomes of research studies as failures. Since the underbelly of the ecosystem is somewhat driven by ``publish or perish'' ideology~\cite{dahlgren_2022}, researchers tend to inherently add significant weight on the number (and quality of venue) of publications. However, we argue (on logical grounds) that publications are means for disseminating new-found results; hence failing to publish in a competitive venue or securing funding should not be considered as failed research. While the scientific community inadvertently uses certain criteria to measure a researcher's success -- citations, publications at high-impact venues, funding acquired -- these may not fully reflect the success of the conducted research. Instead, the criteria reflect how the output is \textit{perceived} by the scientific community in the short term. 
For example, a publication about a specific failure may be successful on its own by preventing others to repeat said failure, hence having a big impact on the research that is performed. 
But this publication has a low chance of being cited in the hopefully successful research that it lead towards.
As such, we first discuss different categories of results obtainable from research studies. 

\subsection{Categories of Research Results} \label{subsec:resultCateg}

\smallskip
\noindent \textbf{I. Positive results} are the results that show the validity of the research hypothesis. Studies with positive results are more likely to be cited, indicating a bias in community in publishing works showcasing positive results \cite{gaillardSystemic2022, nimpfWhy2022}. 

\smallskip
\noindent \textbf{II. Negative results} are the results that contradict previously-established knowledge (e.g., lack of statistical significance to prove the hypothesis of these earlier studies or statistically-significant results contradicting the earlier positive results). 
These negative results, if not emerging due to some flaws in research\footnote{https://www.nature.com/articles/nature.2012.10249}$^{,}$\footnote{https://www.nature.com/articles/d41586-019-01675-9}, can play a significant role in advancing science and avoiding waste of resources (e.g., researchers' time or public money). Hence, before disclosing such results, it is crucial to have confidence that the research is rigorously designed and executed~\cite{bespalov20191312}. 
Additionally, negative results can prove a unique collaboration opportunity for involved research groups to understand the possible reasons for the discrepancy between the obtained results. Negative results are also related to reproducibility. Recreation of previous studies might be affected by uncontrollable configuration changes, design and assumptions, which can lead to different conclusions. Therefore, it is paramount to follow open science practices to facilitate easier investigation of the root causes of discrepancy~\cite{reproducibility}. According to a small-scale survey with 96 researchers found that the majority of the respondents saw value in sharing the negative research outcomes, but do not attempt publishing due to considerable time and effort investment with potentially less number of citations compared to the studies with positive results~\cite{Echevarria2021}. 

\smallskip
\noindent \textbf{III. Unexpected or unexplainable results} are results that cannot be explained with the hypothesis being investigated, requiring further exploration and re-assessment of problem formulation, assumptions, or design choices. Unexpected results can be likely cause of flaws in methodology (e.g. software errors) or unrealistic assumptions, but might also emerge due to the complex system interactions not captured by the hypothesis or designed experiments. For example, in pervasive computing, many factors from the used sensors, their calibration and the composition of study participants play a significant role and can lead to a mismatch between expected and obtained results. 


Not all research studies contain results that can be proved or disproved. \textit{Vision papers} raise important questions pointing to new possibilities, and draw an agenda pioneering a field or vision for researchers to pursue. For opening new avenues and out-of-box thinking beyond the state-of-the-art, vision papers such as~\cite{tennenhouse2007towards} play a crucial role in advancing science. Similarly, \textit{survey studies} and \textit{tutorial papers} can provide a good understanding of the field and assist researchers in identifying the knowledge gaps. 

\subsection{The Cycle of Research} \label{subsec:researchCycle}

\begin{figure}[t]
\centerline{\includegraphics[width=\columnwidth]{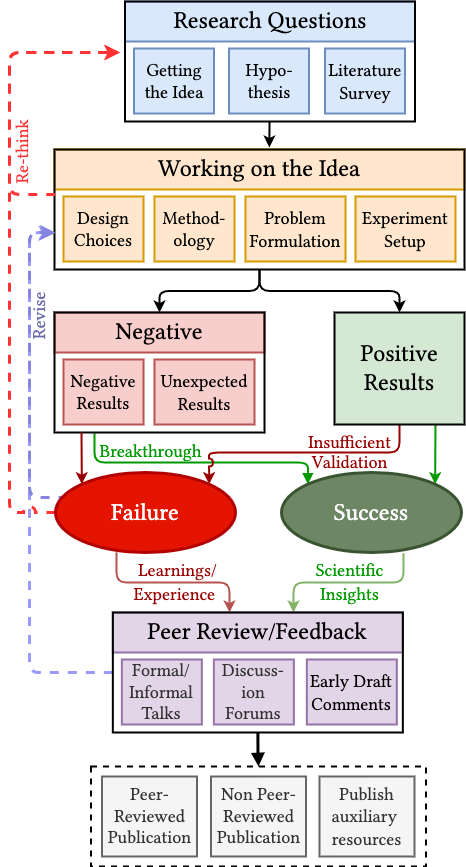}}
\caption{The cycle of research.}
\label{fig:cycle_of_research}
\end{figure}

We now discuss the cycle of research that can potentially lead to such different achievable results from a study, illustrated in Fig.~\ref{fig:cycle_of_research}. To conduct research that advances the field and is impactful, researchers should raise the ``right'' questions to work on. Finding good research questions that fill a knowledge gap may be challenging, as research directions that lie at the intersection of being solvable and pertinent to the community can be limited or require a deep understanding of the literature. At this stage of \textit{formulating the research questions}, apart from performing extensive literature review with an inquisitive mind, researchers can benefit from discussing hypotheses with colleagues (in addition to advisors) or in discussion forums such as Ph.D. schools, workshops, and community meetings. 

Note that performing actions within this stage in conjunction with the next stage of the cycle, i.e., plans to execute the research, is likely to yield desired positive outcomes. For example, suppose the research question involves understanding user behavior over certain stimuli. In that case, it is beneficial to prototype design/methodology for conducting the study, enumerate parameters that need to be recorded and the expected results from the campaign. It is pertinent that researchers are not rigid with their initial assessment of the suitability of the research questions and means to achieve it to avoid \textit{cognitive bias} -- that can significantly impact the outcome of the research. As such, they should be willing to consider preliminary feedback and \emph{re-assess} their approach, formulation and (if required) research direction -- since re-aligning at these initial stages can maximize research impact with the least negative consequences.

Once the research formulation is finalized, researchers should periodically assess the intermediary results from iterative prototypes for revisions, if necessary. Most importantly, it is essential to keep in mind that the experiments can likely yield negative or unexpected results at the end of the study. At this stage, researchers must discover the reason behind their unexpected results -- is it due to wrongful formulation or lack of collected parameters, or is it challenging the widespread assumption within the field? As such, a negative outcome can result in a successful study with a \textit{breakthrough}, albeit a strong validation proves no shortcomings/bias in methodology and problem formulation. Similarly, a positive result from a study does not immediately equate to a successful outcome and requires statistically significant validation to prove no bias in the experiment formulation. 

In the next step of the cycle, the researchers usually formulate their findings into research articles and submit them to peer-review venues. While a significant portion of the scientific community questions the accuracy of the current peer-review systems, often terming it to be ``broken''~\cite{dahlgren_2022}, the core utility of the process is to ensure that the submitted research is free from overlooked biases/errors and incorrect information.

It must be noted that despite widely accepted notions, publications are not the \emph{goal} of a research study but merely means for disseminating the new-found knowledge or experiences gained during the course of the research. Successful research can also produce practical systems and artifacts which do not require traditional peer-review conference/journal publications but open-source code with descriptive documentation, website or even blog post. As a result, researchers should aim to engage with the community in alternate methods (in addition to research publications) to receive a holistic peer review and feedback on improving the research. For example, networking and systems research outcomes can include RFC drafts to be discussed in venues like IETF/IRTF. Researchers can give formal or informal talks at conferences and research institutes and advertise early results in mailing lists, community forums, Slack channels, and so on.

Expert peer review of any form (traditional publication as well as via alternate means) must be considered with due diligence, which might uncover earlier overlooked sources of errors in the research methodology. As a result, the researchers should be open to \emph{revise} their research methodology and formulation to address correctness concerns.

Despite repeated improvements, it might be possible for the manuscript to be repeatedly rejected from multiple publication venues for various reasons. The phenomenon is often misrepresented as a ``failed research". However, as previously stated, failing to publish a paper at the desired venue does not necessarily point to insufficient quality of the research study. Different publication venues have different approaches and understanding of the methodology and research approaches; hence, choosing the venue correctly and ensuring that the manuscript is well-written can be crucial. 

Researchers should also consider submitting to less competitive venues to maximize the \textit{impact} and \textit{timeliness} of a research study. For example, a study that facilitates further research with more efficiency~(e.g., via developing a measurement tool) can be as impactful as studies with groundbreaking ideas. Moreover, the impact of many studies can become evident over time with the advances in supporting sciences and technology. It might turn out that peer-review feedback or methodology revisions reveal that research has failed or obtained results are insignificant. Sharing lessons learned from such failed experiments can also help advance the understanding of the field as others can avoid similar pitfalls that impacted the study. However, care must be taken to accurately summarize the sources of errors/learning and potential future directions for improvement which must not be mistaken with publishing incorrect results. If the researchers correct the study's errors, such learnings can be shared in the appendix of the accepted paper for completeness. The researchers can also look for venues that actively publish and discusses failure results catering to their research field, e.g. PerFail, Insights, and JOURNE. Knowing the possible failure points in the research cycle, knowing the right community for the work in question, and writing an article that is acceptable by the relevant community only comes by experiencing research. As a result, learning from failures is one of the key skills supervisors can deliver to their juniors. 

\begin{table*}[h!]
\caption{Summary of potential failure points at different stages of research study.}
\label{tab:failures}
\begin{tabular}{@{}lll@{}}
\toprule
& \textbf{Failure Points} & \textbf{Recommendations}\\ \midrule

\textbf{Problem Formulation} 
    & \begin{tabular}[c]{@{}l@{}}
    Wrong hypothesis\\ 
    Being ahead of time\\ 
    Unclear research questions\\ 
    Low relevance to venue  \\ 
    Lack of  novelty or delta increment to existing works\\ 
    
    \end{tabular} 

    & \begin{tabular}[c]{@{}l@{}}
    Clear demarcation from literature\\ 
    Vision papers\\ 
    Extensive literature review
    \end{tabular}  
    \\ \midrule

\textbf{Study Design and Execution} 
    & \begin{tabular}[c]{@{}l@{}}
    Theory vs. practice \\ 
    Unrealistic assumptions\\ 
    Too simple or too complex abstraction
    \end{tabular}                                                                                               & 
    
    \begin{tabular}[c]{@{}l@{}}
    A clearly-described methodology section\\ 
    Citing relevant resources\\ 
    Detailed model descriptions
    \end{tabular} 
    \\ \midrule

\textbf{Measurements}               
    & \begin{tabular}[c]{@{}l@{}}
    Software/Hardware errors\\
    Human/sensors errors
    \end{tabular}                                                                                       
    
    & \begin{tabular}[c]{@{}l@{}}
    Avoiding ambiguity\\
    Open datasets\\
    Community accepted methodology
    \end{tabular}                     
    \\ \midrule
    
\textbf{Performance Analysis} & Wrong metrics and methodology  & Comparison with state-of-the-art\\ \midrule
\textbf{Publication} & Repeated rejections & Re-assess relevance and publication venue \\ \bottomrule

\end{tabular}
\end{table*} 

\section{Failure Taxonomy}
In this section, we investigate the process of scientific failures. We identify what is a failure and its key characteristics that hint at research outcomes that might fail. Recognizing these specific situations in research might be an early warning sign that a particular research direction might fail. We present methods and preventive actions that can be taken to avoid and mitigate possible failures. The measures include changing the research direction or methodology early on, gathering insights, and publishing the negative results if the research fails at a later stage.

\subsection{The Origin of Failures} 
Research should start with a clear problem statement and a verifiable hypothesis, as a lack of either is a well-known predetermined breaking point for the subsequent research. One of the most critical steps in producing ``\textit{good}'' research is a comprehensive understanding of the problem that is to be solved. If the problem is not understood well enough, it is possible to research something that lies close to a problem and perhaps is an interesting field on its own but fails to produce fruitful results that can be published. While open-ended research in a preliminary phase is useful to collect ideas, it eventually should be fitted into a more solid framework with clearly-defined goals. Experience says that the sooner the failure is identified, the easier it is to fix.

Related work is critical to advance in science, as this step encourages findings on this topic made by other researchers. This could include positive or negative results on a similar topic, to either be a warning of spending time and effort on a problem that is either unsolvable or already solved. Even more important, the time frame of scanning the related work should not be too small, and perhaps 10, 20, or more years ago, a problem was already solved.

%

\subsection{New Research Topics}
It is easy to focus on the wrong part of the problem or misunderstand the significance of the results, either because you are looking at incomplete or incorrect information or trying to interpret different things from the results from what they actually show. Being early on a topic, especially a topic that later becomes \emph{hot}, is what all researchers dream of doing. But being early on a new topic means there is not an established community around, so there might be no general understanding of the significant results or the critical questions in this area. This might seem counter-intuitive because research is about discovering new things, and somebody always has to write the first paper on a new topic. There is an excellent scope of vision papers in such scenarios.

\subsection{Failure Types}
Failures come in all kinds of shapes and forms (see Table \ref{tab:failures}). Following previous overviews of the experiences and lessons learned while conducting research \cite{howell2021Cracks,honig2018undestanding}, we identify the following set of failure types relevant in pervasive computing field:

\begin{itemize}
    \item Unconvincing results which could not be verified because it is not even possible, for example, due to lack of datasets, real-life experiments, or ground truth.
    \item Under or over-performing experiments that could have been possibly fixed, for example, oversights in system design, inadequate or misconfigured infrastructure, or just buggy code.
    \item Setbacks resulting in lessons learned and acquired hindsight, including hypotheses with too limiting or too broad assumptions.
    \item Unconventional, abnormal, or controversial results that contradict community expectations and are thus extremely hard to reliably prove. 
    \item ``Non-publishable'' or ``hard-to-publish'' side-outcomes of the otherwise correctly run studies, including mistrials of experiment methodology or design, preparations for proof-of-correctness of results, and so on. Limited contribution might not be a failure itself, but might become hard to publish as a full research article.
    \item Unexpected roadblocks affecting publications, including ethical concerns, institutional policy breaches, and difficulties in collaboration with commercially-bound parties of the research projects. Ethical and privacy hurdles might only be encountered after a series of measurements have already been conducted, forcing an unwanted halt to the entire research, including any observations made prior to any private data being collected.
\end{itemize}

\section{Failure-Avoidance and Mitigation} 

In previous sections, we have discussed where, when, and how failures in pervasive computing research may occur. Next, we discuss methods and best practices to avoid failures while conducting research. The studies focusing on simulations, performance analysis, and the more technical side of the field are not free from typical problems either. These topics were also covered during the PerFail workshop \cite{berlo2022insights,das2022enabling, heck2022lessons,liaqat2022hindsight}. Some lessons learned discussed in the workshop include:

\begin{itemize}
    \item \textit{Use step-wise study procedures}, failures and setbacks are easier to recognize during the study.
    
    \item \textit{Design for the target group and consult domain experts}, especially if the study concerns human subjects. Seek feedback from domain experts for any modification to the questionnaire.
    
    \item \textit{Prioritize collecting high-quality data} in the first-hand. Consider the quality of labeling the data items and events, especially when collecting the data in the wild. The sensor devices have to ensure the quality of the signal, even though existing noise.
    
    \item \textit{Ensure validity of the ground truth data} by considering ``gold standard'' devices, especially with sound studies evaluating self-gathered field data with smartphones and wearable devices. The ground truth data must be validated in the lab and in-the-wild environments. Data collection can be monitored via a dashboard to filter noise and abnormality.
    
    \item \textit{Avoid ambiguous questionnaires}. If the study includes real people, accommodate your research for the situation. If using questionnaires, create them in a self-explainable way to avoid misunderstandings. Keep studies and questionnaires as short as possible to improve acceptance and participation rates.
    
    \item \textit{Relation between data selection and experimental plans}. Based on the requirements and practical techniques, selecting a fit dataset affects the final results of the research.
    
    \item \textit{Analyze previous methodologies for deep comparison and understanding}. After deciding on a fit dataset, understanding the dataset needs an effort in analyses from earlier works for data processing and methodologies. Understanding the dataset for experiments is to avoid a lack of important features that fit the experimental requirements. 

    \item \textit{Consider resources related to the experiments}. The experimental results mainly indicate the boost in a feature; however, the trade-off from other aspects or resources needs to be carefully considered, especially when transforming methodologies into a new environment.
\end{itemize}

Additionally, a careful consideration of ethical guidelines is essential~\cite{franzke2020ethics}. This concern may not only protect the integrity of subjects of a study but may also prevent the halt of entire studies. To better safeguard any research involving personal data, the use of Institutional Review Boards is strongly advised and even often required by outlets~\cite{Thomas2017}.

Exciting and risky research will necessarily have more failures than successes; this needs to be understood and normalized in scientific publications. Unlike industry and startup projects, not every research project must yield a success story, as even the insight of a methodology not working for a given problem is a revelation on its own. There are some research directions that are both risky and expensive, where a registered experiment publication (e.g., RSOS\footnote{\url{https://royalsocietypublishing.org/journal/rsos}}) or any publication about negative results is vital to prevent such sunk costs from happening again.

In addition, it is notorious that there are unexpected outcomes or a hypothesis that does not hold~\cite{borji2018CV}. In these cases, it is crucial to show the importance of negative results and lessons via parameters, including the design of experiments, evaluation of performance, comparison of models, statistical hypothesis testing, and so on. It needs to be emphasized that there is a difference between negative results and no results; that the no-results indicate incomplete or unreliable works. Meanwhile, the negative results contain valuable information; they should be welcomed  to save efforts in redundant works and contribute intellectual side related to models, methods, and datasets.

\subsection{When to Fix Failures}
During the research work or latest in the review phase, the modifications in the original manuscript are still possible and even desirable for the readability and cohesion of the article. However, identifying the failures after the manuscript has passed the review cycle and got published is not only more than welcome but also crucial for the reliability of the research. Publishing a retraction, errata, or follow-up are all critical tools to avoid just ``living with'' an error that could have been fixed.

Failure should obviously be avoided despite the pressure to publish and approaching deadlines. This makes lessons learned so critical -- learning from one's own experience and the community as a whole. ``Intelligence is the ability to adapt to change'', which also applies to the research plans and how it actually turns out. This means that research work should be refactored in even late stages if preliminary results necessitate this action.

\section{A Call-for-Arms for Healthy Failure Culture in Research}



Even the most experienced researchers can come with a research question and hypothesis that proves wrong. If the outcome is not expected or aimed for, is this a failure? More importantly, how should it be reported in the community that the first hypothesis did not work? Publishing failures results enable others to avoid making the same mistakes, yet there are significantly more non-working solutions than good, efficient ones. 

Most of research is what is commonly called ``incremental'', i.e., it involves small perturbations of previous research. Done correctly, it entails reproducing the earlier results and adding some variation. Even if the variation does not yield the expected results, the work has successfully reproduced the earlier work, and thus has validated that work. In our opinion, this is definitely not a failure. Unfortunately often such ``failed'' variations are reported as successes, by sugarcoating the results or trying to spin them as positive results. Partly this is because we, as a research community, expect good and positive results, so we should be more tolerant and recognize that reproducing earlier results also carries significant value.

Some research aims at creating something radically new and different; in most cases, these end up failing. Some funding agencies, e.g., European Research Council, explicitly call for \textit{high risk, high gain} types of proposals, while others focus (implicitly) on more incremental work with more guaranteed but smaller successes. We should understand that both kinds of research have value and that failures may simply be indicators of the level of ambition. To propagate a healthier failure culture in research, we propose a call-for-arms action in the community (with a broader relevance outside pervasive computing). 


\subsection{``Guardians of Publication Quality''}

The review process itself is not free of criticism. What will be accepted is, by human nature, dependent on the opinion of reviewers, accompanied by the game of chance to get the reviewers assigned that view the publication most positively~\cite{dahlgren_2022}. The same game of chance is present in the unregulated nature of paper reviews, which allows for claims and unfairness by the reviewers if the editors do not secure uniform quality reviews.

The task of a reviewer or editor \emph{should not} be about improving the prestige of the publication venue, but instead checking the correctness of the work - to the extent feasible. Of course, not all problems in the review process are due to ``malicious'' reviewers. The difficulties are real in conveying the applicability of studies regardless of their limited metrics and comparing publications against each other, as the performance metrics applied to decide the fate of one submission can not be used for another.  In addition, there is a constant need for novel openings instead of building on old theories, despising potential but risky new research fields. Highly selective conferences tend to favor ``exciting'' results, leaving little to no room for the ``correct but boring'' work that has been done rigorously yet falls under the not-so-successful variation of earlier work. 

Yet, an interesting question is how much publication space should failures or negative results get? If the result refutes earlier work, it deserves its spot in the limelight. However, if it simply highlights that a particular variation/configuration did not work, how should it be communicated? Should publication venues dedicate some space to ``failures'', dictating authors to report such unexpected results in appendix or a specifically reserved section? Or should we have dedicated publication venues for publishing such negative results? The former allows readers to learn the underbelly evolution of the work alongside positive results, providing a new learning perspective to the research study. On the other hand, the latter would give a general prominence to ``things that did not work'' in various open problems in the field, encouraging people to take more risks, knowing they have a good chance of advancing the state-of-the-art through unexplored approaches.

\subsection{Normalize Failures as Researchers}

Our aim, through this paper, is to normalize failures as part of the research process and motivate their means for communication. Optimally, this would be through encouraging publication venues to integrate lessons learned from negative results as part of their program. Other means, such as social media or reports on arXiv could also serve as a venue for bringing information about negative outcomes to the public. We also encourage the research community to organize workshops, PhD forums, and other early-stage work events for getting feedback and sharing information on how different researchers -- also those in senior positions -- have overcome the setbacks they have faced.

Communicating stories of the research process as a whole should be seen as equally important as communicating the final results. This could be a section at the end of the article, similarly to ``limitations'' section currently included in the discussion of the results. In addition, communicating the story of the work, from the initial hypothesis through failures to the final decision, should become a standard when discussing the results in conferences or sharing results on (research) social media, such as Twitter, ResearchGate, or LinkedIn. Here, encouraging senior researchers as role models to communicate about setbacks and failures -- and how they did or did not overcome them -- is crucial. Some could even consider publishing a list of rejections along with acceptances. Normalizing failure could even encourage researchers to take more risks in research without the fear of not having to show anything.

\section{Conclusion}

Failures and setbacks in research work are natural and will occur for everyone, from junior PhD students to senior professors. In pervasive computing research, reasons for failures can become apparent quite late and might be unrecoverable, including overlooked aspects in data gathering or instrumentation, wrong assumptions, trust, and ethical considerations. Causes for failures can be common but relatively unknown as published research tends to highlight novel findings and overlook negative results, setbacks, and stories of difficulties. Sometimes writing and getting negative results published can be, indeed, much more difficult than publishing a paper with minor incremental results. 

As we have highlighted in this paper, venues for discussing experiences from failures must become more mainstream. Shared experiences and lessons learned will help others to avoid failures; this should be done by presenting setbacks and talking to other members of the community starting from the early stages of the study. It would be exceptionally important to have a platform where people can publish the stories behind successful papers describing the problems they ran into. This could also be a specific section in each paper for failed trials, similarly to how limitations are currently discussed. For our own part, we will continue running the PerFail workshop on negative results in pervasive computing. We encourage increasing the number of similar workshops also in other fields of the computer science.

\section{ACKNOWLEDGMENT}
Authors are thankful for the participants of the First International Workshop on Negative Results in Pervasive Computing (PerFail2022) and the organizers, especially workshop chairs, of the 20th International Conference on Pervasive Computing and Communications (IEEE PerCom 2022). The paper only reflects on opinions of the authors.

\balance
\bibliographystyle{IEEEtran}
\bibliography{References}

\end{document}